\documentclass[superscriptaddress,preprintnumbers,amsmath,amssymb,twocolumn,floats]{revtex4-1}
\usepackage{txfonts}
\usepackage{amssymb}
\usepackage{graphicx}
\usepackage{sidecap}
\usepackage{CJK}
\usepackage{txfonts}
\usepackage{url}
\usepackage{color}
\begin{document}

\title{Photoemission Spectroscopic Evidence for the Dirac Nodal Line in Monoclinic Semimetal SrAs$_3$}

\author{Y. K. Song}
\thanks{Equal contributions}
\affiliation{State Key Laboratory of Functional Materials for Informatics, Shanghai Institute of Microsystem and Information Technology, Chinese Academy of Sciences, Shanghai 200050, China}
\affiliation{School of Physical Science and Technology, ShanghaiTech University, Shanghai 200031, China}
\affiliation{Center of Materials Science and Optoelectronics Engineering, University of Chinese Academy of Sciences, Beijing 100049, China}

\author{G. W. Wang}
\thanks{Equal contributions}
\affiliation{Department of Physics and Center for Advanced Quantum Studies, Beijing Normal University, Beijing 100875, China}

\author{S. C. Li}
\affiliation{National Laboratory of Solid State Microstructures and Department of Physics, Nanjing University, Nanjing 210093, China}

\author{W. L. Liu}
\affiliation{State Key Laboratory of Functional Materials for Informatics, Shanghai Institute of Microsystem and Information Technology, Chinese Academy of Sciences, Shanghai 200050, China}
\affiliation{School of Physical Science and Technology, ShanghaiTech University, Shanghai 200031, China}

\author{X. L. Lu}
\affiliation{State Key Laboratory of Functional Materials for Informatics, Shanghai Institute of Microsystem and Information Technology, Chinese Academy of Sciences, Shanghai 200050, China}
\affiliation{University of Chinese Academy of Sciences, Beijing 100049, China}

\author{Z. T. Liu}
\affiliation{State Key Laboratory of Functional Materials for Informatics, Shanghai Institute of Microsystem and Information Technology, Chinese Academy of Sciences, Shanghai 200050, China}
\affiliation{CAS Center for Excellence in Superconducting Electronics (CENSE), Shanghai 200050, China}

\author{Z. J. Li}
\affiliation{State Key Laboratory of Functional Materials for Informatics, Shanghai Institute of Microsystem and Information Technology, Chinese Academy of Sciences, Shanghai 200050, China}
\affiliation{CAS Center for Excellence in Superconducting Electronics (CENSE), Shanghai 200050, China}

\author{J. S. Wen}
\affiliation{National Laboratory of Solid State Microstructures and Department of Physics, Nanjing University, Nanjing 210093, China}
\affiliation{Collaborative Innovation Center of Advanced Microstructures, Nanjing University, Nanjing 210093, China}

\author{Z. P. Yin}
\email{yinzhiping@bnu.edu.cn}
\affiliation{Department of Physics and Center for Advanced Quantum Studies, Beijing Normal University, Beijing 100875, China}

\author{Z. H. Liu}
\email{lzh17@mail.sim.ac.cn}
\affiliation{State Key Laboratory of Functional Materials for Informatics, Shanghai Institute of Microsystem and Information Technology, Chinese Academy of Sciences, Shanghai 200050, China}
\affiliation{CAS Center for Excellence in Superconducting Electronics (CENSE), Shanghai 200050, China}

\author{D. W. Shen}
\email{dwshen@mail.sim.ac.cn}
\affiliation{State Key Laboratory of Functional Materials for Informatics, Shanghai Institute of Microsystem and Information Technology, Chinese Academy of Sciences, Shanghai 200050, China}
\affiliation{Center of Materials Science and Optoelectronics Engineering, University of Chinese Academy of Sciences, Beijing 100049, China}
\affiliation{CAS Center for Excellence in Superconducting Electronics (CENSE), Shanghai 200050, China}

\begin{abstract}
Topological nodal-line semimetals with exotic quantum properties are characterized by symmetry-protected line-contact bulk band crossings in the momentum space. However, in most of identified topological nodal-line compounds, these topological non-trivial nodal lines are enclosed by complicated topological trivial states at the Fermi energy ($E_F$), which would perplex their identification and hinder further applications. Utilizing angle-resolved photoemission spectroscopy and first-principles calculations, we provide compelling evidence for the existence of Dirac nodal-line fermions in the monoclinic semimetal SrAs$_3$, which are close to $E_F$ and away from distraction of complex trivial Fermi surfaces or surface states. Our calculation indicates that two bands with opposite parity are inverted around \emph{Y} near $E_F$, which results in the single nodal loop at the $\Gamma$-\emph{Y}-\emph{S} plane with a negligible spin-orbit coupling effect. We track these band crossings and then unambiguously identify the complete nodal loop quantitatively, which provides a critical experimental support to the prediction of nodal-line fermions in the CaP$_3$ family of materials. Hosting simple topological non-trivial bulk electronic states around $E_F$ and no interfering with surface states on the natural cleavage plane, SrAs$_3$ is expected to be a potential platform for topological quantum state investigation and applications.
\end{abstract}

\maketitle


Topological semimetal, as the non-trivial extension of topological classification of electronic quantum states from the insulator to metal, is a group of materials in which the conduction and valence bands cross and form nodes behaving as monopole of a Berry flux \cite{1,2}. When nodes are close to the Fermi energy ($E_F$), the low-energy quasiparticle excitation would be drastically different from that of the conventional Schr\"{o}dinger-type fermion and thus lead to novel transport properties, which are crucial to the further study of novel quantum states and modern quantum devices \cite{3,4}. While some distinct point-contact nodes, e.g., 4-, 2-, and 3-fold degenerate nodes, have been confirmed in Dirac \cite{5,6}, Weyl \cite{7,8,9,10,11,12,13,14}, and triply-degenerate semimetals \cite{15,16} in previous studies, respectively, line-contact nodes, i.e., nodal lines, with their various configurations \cite{17} have not been fully investigated in experiments until now.

The nodal ring, nodal link and nodal chain all belong to nodal-line systems in which nodes extend along one-dimensional lines instead of discrete points in the three-dimensional (3D) Brillouin zone (BZ) \cite{3,17,18,19,20,21,22,23,24,25,26,27,28,29,30,31,32,33,34}. Theories have predicted that a non-trivial Berry phase around the nodal line would generate a half-integer shift of Landau-level index \cite{20} and result in drumhead-like surface states \cite{19,21}. To date, nodal-line states have been theoretically proposed and then experimentally confirmed in several compounds, including CaAgX (X=P, As) \cite{22,23}, PbTaSe$_2$ \cite{24}, ZrSiS \cite{25,26,27,28}, and MB$_2$ (M=Ti, Zr) \cite{29,30,31,32}. However, for most of these compounds, topological non-trivial band-crossing nodes are enclosed by complex trivial bulk or surfaces bands, which would complicate potential applications of these topological nodal-line semimetals.

\begin{figure}[t]
\includegraphics[width=86mm]{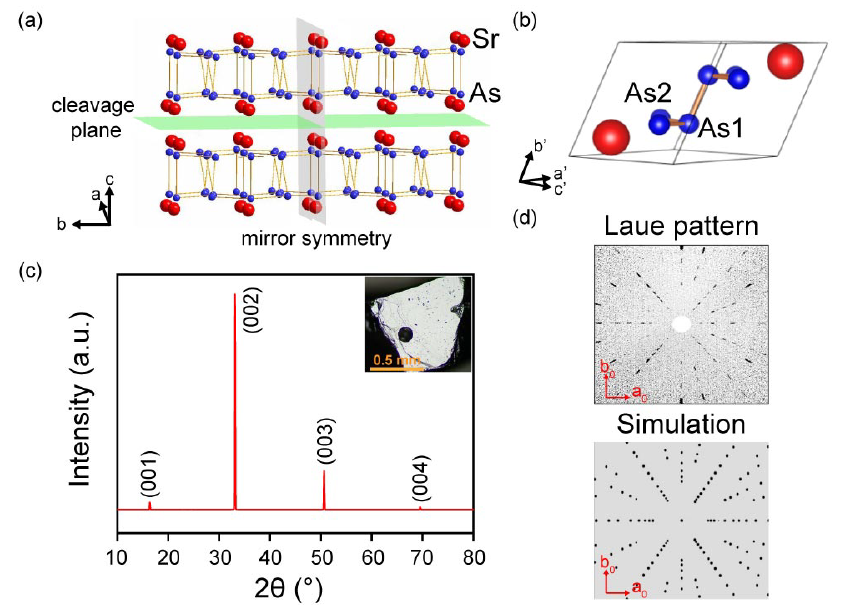}
\caption{(a) The side view of the crystal structure of SrAs$_3$. The atomic layers of As and Sr alternately stack along the \emph{c} axis, and the natural cleavage and mirror symmetry planes are represented by the green and gray planes, respectively. (b) The monoclinic primitive cell of SrAs$_3$. (c) The single-crystal XRD measured on the (001) plane of the conventional lattice of SrAs$_3$. The inset shows the flat and shining cleavage plane. (d) The comparison between the X-ray Laue pattern taken on the cleaved sample (upper) and simulation (down).}
\label{crystal structure}
\end{figure}

\begin{figure*}
\includegraphics[width=15cm]{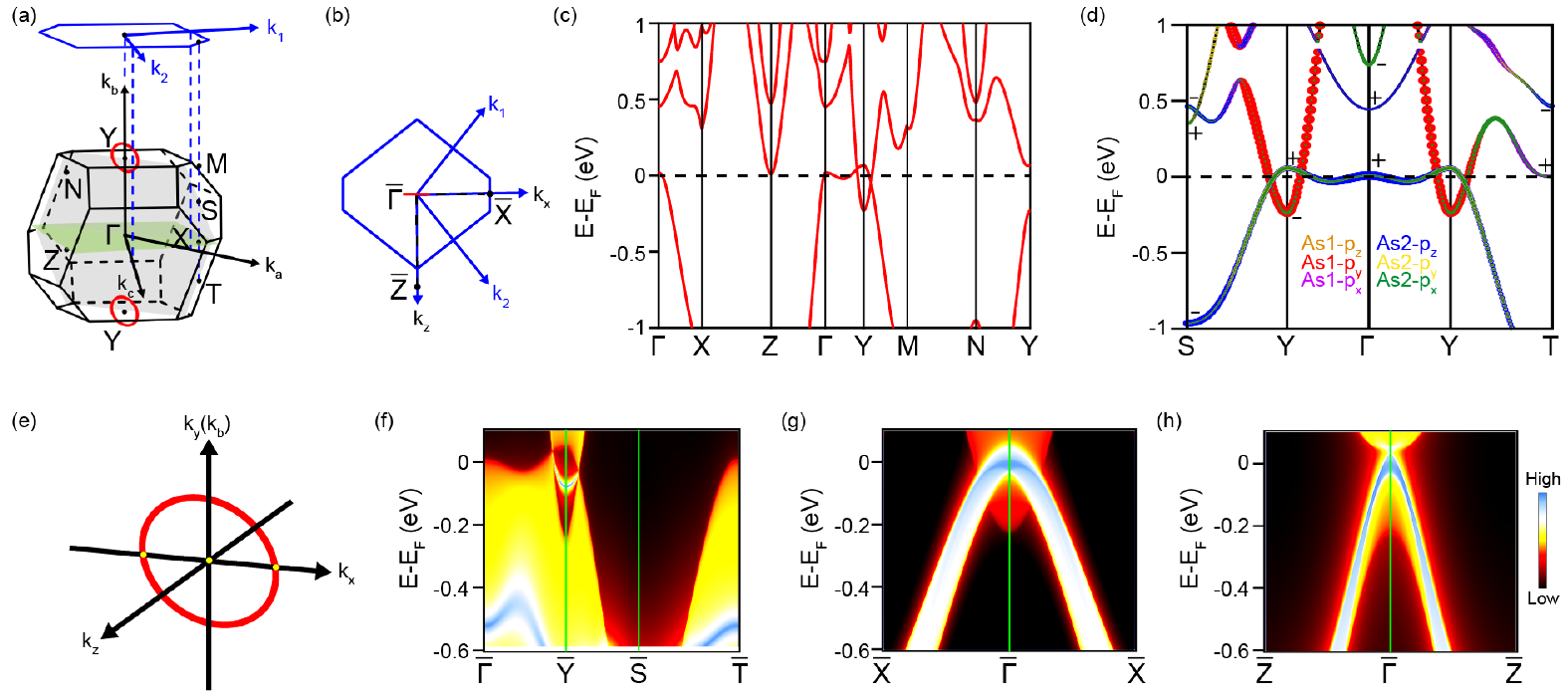}
\caption{(a) The 3D BZ of the SrAs$_3$ primitive lattice. The nodal loop (red ring) surrounds the \emph{Y} point on the $\Gamma$-\emph{Y}-\emph{S} plane, i.e., the mirror symmetry plane. The green plane is perpendicular to the \emph{k$_y$} (\emph{k$_y$}) axis. (b) The projected BZ perpendicular to \emph{k$_b$}. The red line is the projection of the nodal loop. $k_1$ and $k_2$ represent projections of $k_a$ and $k_c$ onto this plane, respectively. (c) The calculated band structure along high-symmetry paths without SOC. (d) The calculated band structure with orbital characters and parity analysis. (e) The schematic of the nodal loop on the \emph{k$_x$}-\emph{k$_y$} plane. (f) The calculated surface state in the projected BZ along the $k_c$ direction. The drumhead-like surface state is nestled between two solid Dirac cones, which are the projection of the nodal loop. (g), (h) Calculated band structure along different high-symmetry paths in the project BZ along the $k_y$ direction.}
\label{bands}
\end{figure*}

Very recently, the CaP$_3$ family of materials have been predicted to possess topological nodal rings with two-dimensional drumhead-like surface states, which are protected by the spatial-inversion and time-reversal symmetries~\cite{33,34}. Specifically, for the monoclinic semimetal SrAs$_3$, a nodal loop was proposed to be located at $E_F$, without the distraction of complicated trivial topological Fermi surfaces (FSs). If the spin-orbit coupling (SOC) is considered, a small gap would open along the loop, resulting in a strong topological insulator~\cite{34}. Actually, magnetoresistance (MR) and Shubnikov-de Haas quantum oscillation measurements on the single-crystal SrAs$_3$ have revealed the negative MR induced by the chiral anomaly and the non-trivial $\pi$ Berry phase, which strongly suggest the existence of topological nodal-line semimetal states therein~\cite{38,39}. However, so far  direct spectroscopic evidence of such novel band structure on that is still lack.

In this Letter, we provide  compelling evidence for Dirac nodal-line fermions in the monoclinic semimetal SrAs$_3$ by means of angle-resolved photoemission spectroscopy (ARPES) and first-principles calculations. Our calculations suggest that a single nodal loop should be located on the $\Gamma$-\emph{Y}-\emph{S} plane, which would not be distracted by complex topological trivial FSs. Further detailed photon-energy-dependent ARPES measurements on samples with naturally cleaved (010) planes unambiguously reveal the complete nodal-line feature around the \emph{Y} point, which well reproduces calculations along the corresponding paths in the BZ and then confirms the existence of nodal-line fermions in the CaP$_3$ family definitely.

High-quality single crystals of SrAs$_3$ were grown by solid-phase sintering reaction using stoichiometric amounts of Sr and As as described elsewhere~\cite{38}. ARPES measurements were performed at 13U beamline of National Synchrotron Radiation Laboratory (NSRL), Dreamline and 03U beamlines of Shanghai Synchrotron Radiation Facility (SSRF), and One-Squared ARPES endstation at BESSY II. The energy and angular resolutions were set to around 10 meV and 0.2$^\circ$, respectively. All samples were cleaved under a vacuum
better than $5\times10^{11}$ Torr. During measurements, the temperature was kept at 20 K. First-principles calculations were performed by using the linearized augmented plane wave method as implemented in WIEN2K \cite{40} combined with Perdew-Burke-Ernzerhof form of the general gradient approximation to the exchange-correlation functional \cite{41}. Additionally, the modified Becke-Johnson (MBJ) potential \cite{42} was employed for a reasonable correction, as GGA may give underestimated relative positioning of valance and conduction bands for semiconductors and semimetals. In order to explore the topological properties, a tight-binding model based on the maximally localized Wannier functions (MLWF) was constructed using the WANNIER90 \cite{43,44} and WIEN2WANNIER \cite{45} programs from the density functional theory (DFT) band structures. The surface spectral functions were calculated by the iterative Green’s function method implemented in the WANNIERTOOS code \cite{46}.

The crystal structure of SrAs$_3$ belongs to the simple monoclinic space group [$C2/m$ (No. 12)], in which Sr layers form channels in the \emph{a}-\emph{b} plane while As atoms are inserted into these channels, as illustrated in Fig. 1(a). It can be viewed as a list of infinite atomic layers of As and Sr alternately stacking along the $c$ axis. In this way, the natural cleavage plane should be located in between the neighbouring Sr layers [the green plane shown in Fig. 1(a)]. The lattice is reflection symmetric with respect to the -Sr-As-As-Sr- atomic plane marked by gray plane, and this crystallographic symmetry provides additional protection for the topological non-trivial electronic structure. Moreover, for this $C2/m$ space group compound, its primitive unit cell is shown in Fig. 1(b). It contains two types of As atoms, with As1 on the mirror plane and As2 on both sides.

Considering the sensitivity of electronic structure to the crystal structure and stoichiometry, we carefully characterized our SrAs$_3$ samples. We confirmed that these samples are pure-phase SrAs$_3$ with the molar ratio of Sr:As = 1:3 by using both the powder X-ray diffraction (XRD) and scanning electron microscope equipped with an energy dispersive X-ray spectrometer (EDS) (see more details in Supplemental Material \cite{47}). After cleaved in the air, our samples show typical flat and shining surfaces [the inset of Fig. 1(c)], and these samples have been further characterized by the single-crystal XRD, as shown in Fig. 1(c). The results reveal their monoclinic crystal structure with the deduced lattice constants a=9.605(7) {\AA}, b=7.665(3) {\AA}, and c= 5.873(1) {\AA}, and $\alpha$ = $\gamma$ = 90$^\circ$, $\beta$=112.869$^\circ$, in an excellent agreement with the previous report \cite{48}. The single-crystal XRD measurement as well indicates that the \emph{a}-\emph{b} plane should be the natural cleavage plane stacking up along \emph{c} axis. Moreover, as shown in Fig. 1(d), the sharp X-ray Laue pattern taken from the cleaved sample can be well reproduced by the simulated result on the \emph{a}-\emph{b} plane, which further confirms the cleavage plane and high quality of our samples. We note that the naturally cleaved surface of SrAs$_3$ actually corresponds to the (010) plane of its primitive unit cell. For simplicity, hereafter we will always apply the primitive unit cell in our discussion.

Fig. 2(a) illustrates the 3D BZ of the SrAs$_3$ primitive lattice, in which the gray plane (the $\Gamma$-\emph{Y}-\emph{S} plane) stands for the mirror symmetry plane. According to the calculated band structure along high-symmetry lines [Fig. 2(c)], one can see that the band structure of SrAs$_3$ around \emph{E$_F$} is rather simple. Except for a hole-like band barely crossing \emph{E$_F$} around $\Gamma$, there is only one band inverted structure around the \emph{Y} point. Actually, our calculations further show that these two bands bear opposite parity and invert in the Fermi energy only around \emph{Y} in the whole BZ, as shown in Fig. 2(d). We as well performed the orbital character analysis on bands in the vicinity of \emph{E$_F$}, and they are mainly derived from $p$ orbitals of As atoms. Specifically, the conduction band around the \emph{Y} point is dominated by As1-\emph{p$_y$} orbital, while the valence band is mainly contributed by As2-\emph{p$_z$}, As2-\emph{p$_y$} and As2-\emph{p$_x$} orbitals [Fig. 2(d)]. Theoretically, with both the spatial-inversion and time-reversal symmetries reserved, the energy inverted bands with opposite parity have been proposed to cross along a closed nodal loop~\cite{34}. In the case of SrAs$_3$, calculations suggest that a closed Dirac nodal loop should surround the \emph{Y} point on the $\Gamma$-\emph{Y}-\emph{S} plane, just as the red circle marked in Fig. 2(a). As illustrated in Fig. 2(f), our semi-infinite slab calculations clearly resolve the characteristic drumhead-like surface state around $\overline{Y}$ in the projected BZ along the $k_c$ direction, which is nestled between two Dirac cones projected by the nodal loop in this direction. However, in the projected BZ along $k_y$($k_b$), which is actually reachable by ARPES on cleaved SrAs$_3$, the small drumhead-like surface state is invisible since it is buried in bulk states [See Figs. 2(g-h)]. Besides, even when the SOC is considered, the nodes would open a gap much smaller than those of other previously reported nodal-line semimetals, suggesting the insignificant SOC effect on its bulk electronic structure (see more details in Supplemental Material \cite{47}). In these regards, ARPES measurements on (010) cleavage plane of SrAs$_3$ would provide us a unique opportunity to investigate the topological nodal line without the interference of other complicated bulk or surface states in the vicinity of \emph{E$_F$}.

\begin{figure*}[htbp]
\includegraphics[width=17cm]{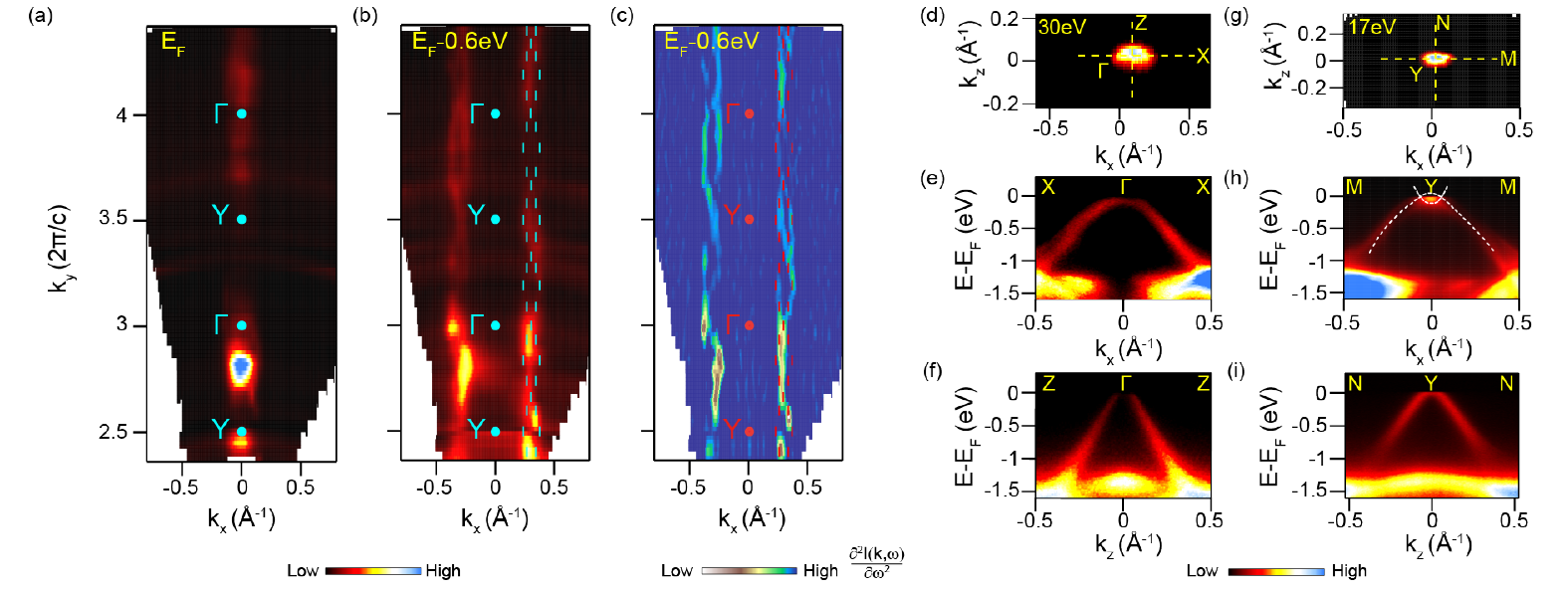}
\caption{(a), (b) Integrated photoemission intensity maps in the \emph{k$_x$}-\emph{k$_y$} plane taken at $E_F$ and $E_F$-0.6 eV, respectively. (c) The corresponding second derivative plot of (b). (d) and (g) Integrated photoemission intensity maps in the $\Gamma$-\emph{X}-\emph{Z} ($hv$=30 eV) and \emph{Y}-\emph{M}-\emph{N} ($hv$=17 eV) high-symmetry plane, respectively. (e), (f) Photoemission intensity plots taken along $\Gamma$-\emph{X} and $\Gamma$-\emph{Z}, respectively. (h), (i) Photoemission intensity plots taken along \emph{Y}-\emph{M} and \emph{Y}-\emph{N}, respectively. }
\label{ky}
\end{figure*}

Since the predicted nodal loop in SrAs$_3$ is exactly perpendicular to the (010) cleavage plane, as shown in Figs. 2(a) and 2(b), a detailed photon-energy-dependent ARPES measurement on \emph{k$_x$}-\emph{k$_y$} plane is highly desired to identify the whole nodal loop [Fig. 2(e)]. Figs. 3(a) and 3(b) show photoemission intensity maps in the $\Gamma$-\emph{X}-\emph{Y}-\emph{M} plane (the \emph{k$_x$}-\emph{k$_y$} plane) taken at $E_F$ and $E_F$-0.6 eV, respectively. Together with the corresponding second derivative plot, one can clearly distinguish the warped chain-like band feature with a pronounced periodic modulation along the \emph{k$_y$} direction, as shown in Fig. 3(c). An inner potential of 19.3 eV has been obtained through the best fit to the periodic variation according to the free-electron final-state model~\cite{49}, allowing us to determine photon energies corresponding to high-symmetry planes.

Figs. 3(d)-3(i) show FSs and photoemission intensity plots taken with 30 eV and 17 eV photons, which are corresponding to band structure in the $\Gamma$-\emph{X}-\emph{Z} and \emph{Y}-\emph{M}-\emph{N} high-symmetry planes, respectively. In Fig. 3(d), the FS on the $\Gamma$-\emph{X}-\emph{Z} plane is an elliptical hole pocket, and band dispersions along the $\Gamma$-\emph{X} and $\Gamma$-\emph{Z} directions exhibit anisotropic structure, as displayed in Figs. 3(e) and 3(f), respectively. As for the \emph{Y}-\emph{M}-\emph{N} plane, the FS and band structure are rather similar to those on the $\Gamma$-\emph{X}-\emph{Z} plane, as shown in Figs. 3(g)-3(i). The main difference of band structure between these two high-symmetry planes lies in the existence of an additional inverted band structure producing nodes around the \emph{Y} point, as highlighted by crossing white lines in Fig. 3(h). Note that such band inversion can be only observed along \emph{Y}-\emph{M} direction but is totally absent along \emph{Y}-\emph{N}, which is in a good agreement with our calculations in Figs. 2(g) and 2(h). Actually, around $Y$, both cuts along the \emph{Y}-\emph{M} ($k_x$) and \emph{Y}-\emph{N} ($k_z$) directions would pass through the center of the nodal loop. However, the \emph{Y}-\emph{M} cut would have two intersections with the loop, while the \emph{Y}-\emph{N} should have no intersection with the loop, as illustrated in Fig. 2(e). Moreover, for the band crossings occurring along \emph{Y}-\emph{M}, we did not resolve any visible lifting of the node degeneracy, indicating the actually negligible SOC effect in this compound.

In order to determine the complete node-line feature, we carried out a high-resolution ARPES survey around $Y$ in the $k_x$-$k_y$ plane with varying photon energies, just as illustrated in the inset of Fig. 4(a). When the photon energy is changed, the cut would move almost vertically up and down along the $k_y$ direction. As shown in Fig. 4(a), when the ARPES cut exactly passes the the \emph{Y} point (taken at 17 eV), the electron-like and hole-like bands cross, forming two nodes without interfacing with surface states, which is in a remarkable agreement with calculations (white lines). While, as ARPES cuts have gradually been away from the \emph{Y} point, the band-crossing area would shrink and finally disappear. Such an evolution of this elliptical nodal loop in SrAs$_3$ with the change of $k_y$ has been clearly demonstrated. Here, both our experimental (blue solid dots) and theoretical results (the red line) can profile the complete nodal loop in SrAs$_3$. Besides, the corresponding energy-distribution curves (EDCs) are shown in Fig. 4(b), which could further confirm band-crossing introduced nodes and their evolution with $k_y$ positions. In this way, moreover, we could quantitatively determine the real size of the nodal loop, as marked by blue dots in Fig. 4(c). We found that the height and width of nodal loop along $k_y$ and $k_x$ directions are 0.2$\pm$0.02 ${\AA}^{-1}$  and 0.15$\pm$0.03 ${\AA}^{-1}$, respectively, which are in a good agreement with the prediction. Thus, we have unambiguously tracked the topological non-trivial nodal line in this prototypical topological semimetal.

\begin{figure}[htbp]
\includegraphics[width=9cm]{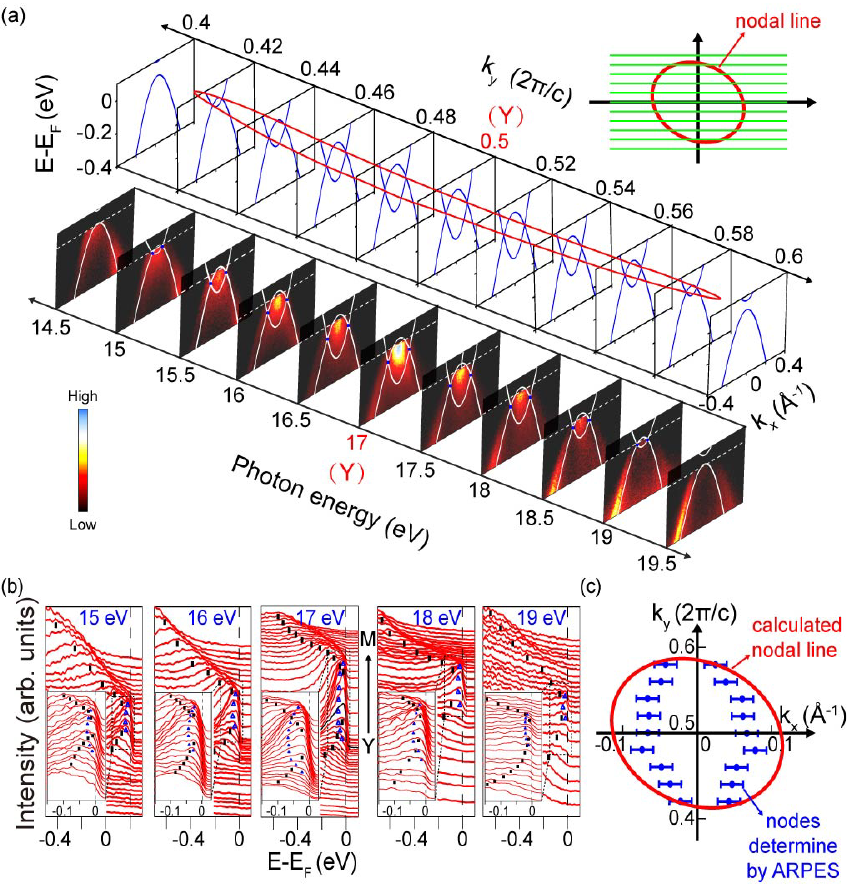}
\caption{(a) Band dispersions along the $\overline{Y}-\overline{M}$ direction taken with different photon energies, and the calculations are illustrated by white lines for comparison. These blue dots represent nodes. The insert shows the schematic of ARPES cuts crossing the nodal loop along $k_y$ on the \emph{k$_x$}-\emph{k$_y$} plane. Green lines represent ARPES cuts at different \emph{k$_y$} positions, which are taken with different photon energies. The red ring stands for the nodal loop. (b) Corresponding EDCs taken at different photon energies. (c) The comparison between nodal loops determined by ARPES (blue dots) and calculations (the red ring). }
\label{nodal line}
\end{figure}


In summary, by using ARPES combined with first-principles calculations, we unambiguously demonstrate the existence of Dirac nodal-line fermions in SrAs$_3$, which are under the protection of the combination of the spatial-inversion and time-reversal symmetries in this $C2/m$ symmetry material. With the insignificant SOC effect and no interfering with topological trivial complicate states around $E_F$, SrAs$_3$ provides a good platform for further studies into Dirac nodal-line fermions.

We acknowledge Dr. L. L. Wang and Dr. H. Jin for their helpful discussion. This work was supported by the National Key R\&D Program of the MOST of China (Grant Nos. 2016YFA0300204 and 2016YFA0302300), the National Science Foundation of China (Grant Nos. 11574337, 11704394, 11674030 and 11822405), and the Fundamental Research Funds for the Central Universities (Grant No. 310421113). Part of this research used Beamline 03U of the Shanghai Synchron Radiation Facility, which is supported by ME2 project under contract No. 11227902 from National Natural Science Foundation of China. D.W.S. is supported by ``Award for Outstanding Member in Youth Innovation Promotion Association CAS''. The calculations used high performance computing clusters of Beijing Normal University in Zhuhai and the National Supercomputer Center in Guangzhou.

\emph{Note added}. We have been aware of a recent ARPES work on SrAs$_3$ with different results with ours\cite{50}. Their samples were grown by Sn-self flux technique, while ours were grown by solid-phase sintering reaction method without any fluxes. We stress that our samples have been strictly characterized by detailed XRD and EDS to confirm their pure phase and right stoichiometry.

\bibliographystyle{apsrev4-1}

%

\end{document}